\documentclass[aps,prb,twocolumn]{revtex4}
\usepackage{amsfonts}
\usepackage{amsmath}
\usepackage{graphicx}
\usepackage{dsfont}
\usepackage{diagbox}
\usepackage{array}
\usepackage{amssymb}
\usepackage{bbm}
\usepackage{float}
\usepackage{url}
\usepackage{hyperref}
\usepackage{booktabs}
\usepackage{xcolor}

\begin{document}

\title{Tensor network approach to the two-dimensional fully frustrated XY model and a chiral ordered phase}
\author{Feng-Feng Song$^{1}$ and Guang-Ming Zhang$^{1,2}$}
\email{gmzhang@tsinghua.edu.cn}
\affiliation{$^{1}$State Key Laboratory of Low-Dimensional Quantum Physics and Department
of Physics, Tsinghua University, Beijing 100084, China. \\
$^{2}$Frontier Science Center for Quantum Information, Beijing 100084, China.}
\date{\today }

\begin{abstract}
A general framework is proposed to solve the two-dimensional fully frustrated XY
model for the Josephson junction arrays in a perpendicular magnetic field. The essential idea is to encode the ground-state local rules induced by frustrations in the local tensors of the partition function. The partition function is then expressed in terms of a product of one-dimensional transfer matrix operator, whose eigen-equation can be solved by an algorithm of matrix product states rigorously. The singularity of the entanglement entropy for the one-dimensional quantum analogue provides a stringent criterion to distinguish various phase transitions without
identifying any order parameter a prior. Two very close phase transitions
are determined at $T_{c1}\approx 0.4459$ and $T_{c2}\approx 0.4532$, respectively.
The former corresponding to a Berezinskii-Kosterlitz-Thouless phase transition
describing the phase coherence of XY spins, and the latter is an Ising-like
continuous phase transition below which a chirality order with spontaneously broken $Z_2$ symmetry is established.
\end{abstract}

\maketitle

\section{Introduction}

It is well-known that the Berezinskii-Kosterlitz-Thouless (BKT) mechanism%
\cite{Berezinsky1970,Kosterlitz1973} provides a prototypical example of
topological phase transitions in two-dimensional (2D) systems and has been
extensively investigated in various systems. The phase coherence of Cooper pairs 
in 2D superconductivity can be characterized by the BKT transition, corresponding to
the unbinding vortices and anti-vortices. One of the prototype models is the 
two-dimensional XY model, and an attractive platform to realize the XY model is Josephson 
junction arrays\cite%
{PRL_47_1542,PRB_26_5268,PRB_35_7291,Fazio_2001,Newrock_2000,Nakamura_2020},
where the XY spin variables represent the superconducting order-parameter phases.
When applying a perpendicular magnetic field such that the flux density per
plaquette is just one-half flux\cite{Zant1991,Ling1996,Martinoli2000,Cosmic2020},
we have the so-called fully frustrated XY model (FFXY).

The FFXY model was proposed originally as a continuum version of spin
glasses possessing competing ferromagnetic and antiferromagnetic interactions%
\cite{Villain1977, Villain1977_2}. Although the model is $U(1)$ invariant, a
new $Z_{2}$ degree of freedom emerges as a result of minimization of local conflict
interactions. Due to the presence of strong frustration, extensive studies has been
carried out for the FFXY model on the square lattice\cite%
{Teitel1983,Thijssen1990,Santiago1992,Granato1993,LeeJR1994,LeeSy1994,Santiago1994,
Olsson1995,Cataudella1996,Olsson1997,Boubcheur1998,Hasenbusch2005,Okumura2011,Nussinov_2014,Lima2019}
or the antiferromagnetic XY spin model on the triangular lattice\cite%
{Miyashita1984,Shih1984,DHLee1984,DHLee1986,Korshunov1986,Himbergen1986,Xu1996,LeeSy1998,Luca1998}. The nature of the phase transitions in the 2D FFXY has been the subject of
a long controversy, because two distinct types of ordering occur extremely
closed to each other\cite{Teitel1983}. Despite much effort has been
dedicated to the study of this model, there is not yet a general consensus
on the critical behavior of these systems\cite{Okumura2011,Lima2019}.

Due to the ground state degeneracies, the study of strongly correlated
statistical systems with frustrations has proven to be really difficult and
most sampling methods suffer from a critical slowing down when approaching
the low temperature phase\cite{Wolff1989}. Recently, an increasing interest
has been stimulated on the investigation of tensor network methods in the
study of the frustrated systems. It is found that, although the tensor
network methods provide an effective computational approach to study the
classical lattice models\cite{Verstraete2008,Orus2014}, special attention
should be paid in the construction of the local tensors in the presence of
geometrical frustrations, which has been demonstrated in the simulations of
frustrated classical spin systems with discrete degrees of freedom\cite%
{Vanderstraeten2018,Vanhecke2021}. The key point is that the ground state
local rules induced by frustrations should be encoded in the local tensors
when comprising the whole tensor network of the partition function.

In this work, we apply the state-of-art tensor network method to study such
strongly frustrated spin systems in the thermodynamic limit. It is
demonstrated that the extension of the applicability of tensor networks to
the fully frustrated systems with continuous $U(1)$ degrees of freedom is
nontrivial, because the standard formulation of the tensor network fails to
converge. Here we propose a new construction strategy based on the splitting
of $U(1)$ spins, and then the partition function of the FFXY model is
transformed into an infinite 2D tensor network with an enlarged unit cell,
which can be efficiently contracted by a recently proposed tensor network
algorithm\cite{Nietner2020} under optimal variational principles\cite%
{Stauber2018,Vanderstraeten2019}.

\begin{figure}[tbp]
\centering
\includegraphics[width=\linewidth]{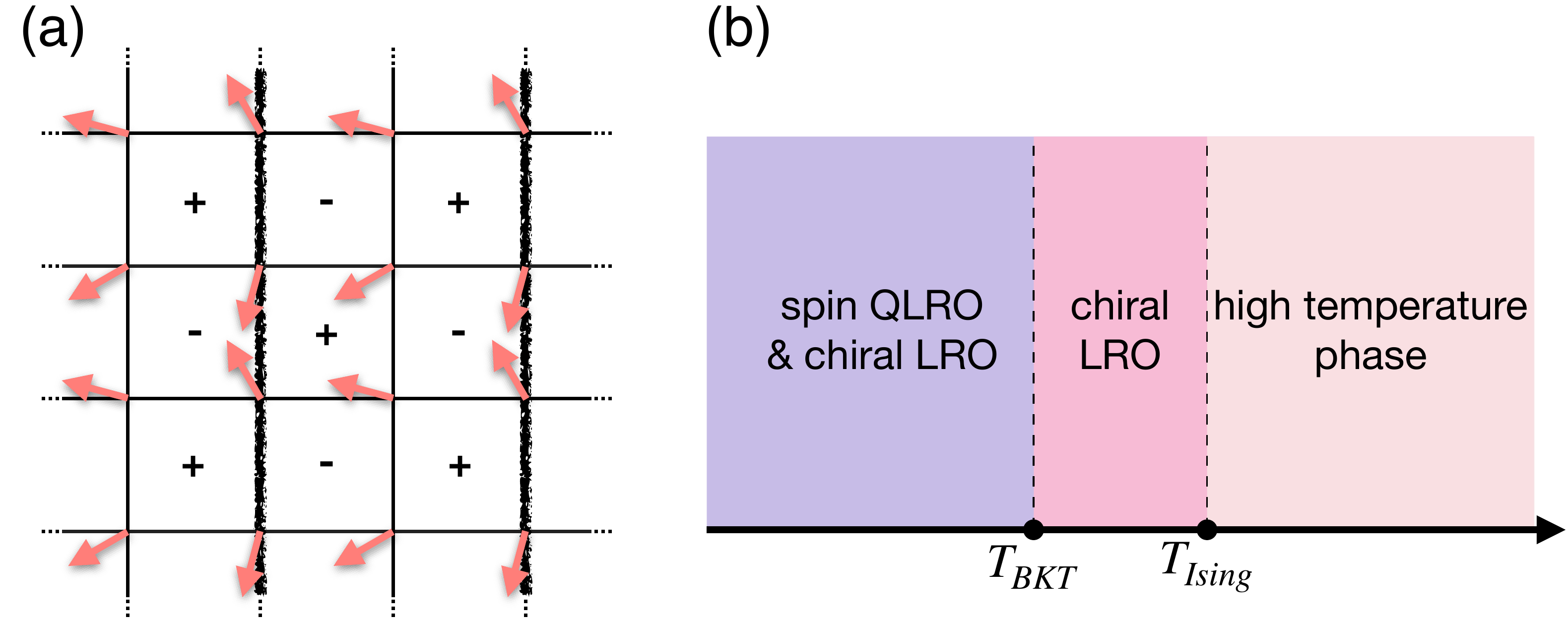}
\caption{ (a) The ground state of the fully frustrated XY model on the
square lattice. The $\pm$ signs denote the checkerboard like chiralities.
Thick lines correspond to $A_{ij}=\protect\pi$ and thin lines to $A_{ij}=0$.
(b) The finite-temperature phase diagram of the FFXY model showing a low
temperature phase with XY spin quasi-long-range order (QLRO) and chiral
long-range order (LRO), a chiral ordered phase with XY spin disorder, and a
high temperature disordered phase. Two different transitions happen at $%
T_{BKT}\simeq0.4459$ and $T_{Ising}\simeq0.4459$ belonging to the BKT and
Ising universality class, respectively.}
\label{fig:phase}
\end{figure}

As the partition function is written in terms of a product of 1D transfer
matrix operator, the singularity of the entanglement entropy of this 1D
quantum transfer operator can be used to determine various phase transitions
with great accuracy\cite{Haegeman2017}. The distinct advantage of the tensor
network method over the Monte Carlo simulations is that a stringent
criterion can be used to distinguish various phase transitions without
identifying any order parameter a prior. From the perspective of the quantum
entanglement, we can resolve the puzzles about the FFXY model with a clear
evidence that the Ising phase transition takes place at a higher temperature
$T_{Ising}$ than the BKT transition $T_{BKT}$. The finite-temperature phase
diagram is displayed in Fig.~\ref{fig:phase}(b). The low temperature phase is characterized by a long-range ordered checkerboard pattern of chirality together with a quasi-long-range XY spin order. In the intermediate temperature region ($T_{BKT}<T<T_{Ising}$), the long-range chiral order survives while the spin-spin correlations are destroyed.

The paper is organized as follows. In Sec. II we give an introduction of the 2D FFXY model and the possible phase transitions. In Sec. III we
develop a general framework of the tensor network theory for this fully frustrated
model. In Sec. IV we present the numerical results for the determination of the
finite temperature phase diagram. Finally in Sec. V, we discuss the nature of the
intermediate temperature phase and give our conclusions.

\section{Fully frustrated XY model}

The fully frustrated XY model on a 2D square lattice can be defined by the Hamiltonian
\begin{equation}
H=-J\sum_{\langle ij\rangle }\cos \left( \theta _{i}-\theta
_{j}+A_{ij}\right) ,  \label{eq:ffxy}
\end{equation}
to describe the Josephson junction array under an external magnetic field\cite{Teitel1983, Zant1991}, where $J>0$ is the coupling strength, $i$ and $j$
enumerate the lattice sites, and the summation is over the pairs of the
nearest neighbors. The frustration is induced by the gauge field defined on
the lattice bond satisfying $A_{ij}=-A_{ji}$. The gauge field is related to
the vector potential of the external magnetic field by $A_{ij}=\frac{2\pi }{\Phi _{0}}\int_{\vec{r}_{i}}^{%
\vec{r}_{j}}d\vec{l}\cdot \vec{A}$, where $\Phi _{0}=\frac{hc}{2e}$ is the
flux quantum. The case of full frustration corresponds to the $Z_{2}$ gauge
field (half quantum flux per plaquette), i.e., $\sum_{\langle i,j\rangle \in
\Box }A_{ij}=\pi $, where the sum is taken around the perimeter of a
plaquette.

The ground states of the FFXY model on a square lattice presents a $%
U(1)\times Z_{2}$ degeneracy\cite{Villain1977,Villain1977_2}. The $U(1)$
degeneracy is related to the global invariance of the Hamiltonian like the
2D XY model. The two-fold discrete degeneracy is resulted from the $Z_{2}$
symmetry of the Hamiltonian under the simultaneous reversal in the signs of
all $\theta _{i}$ and $A_{ij}$. The ground state is characterized by a
checkerboard pattern of chiralities in analogy to the antiferromagnetic
Ising model, where the planar spins rotate clockwise and counterclockwise
alternatively around the plaquettes. As shown in Fig.~\ref{fig:phase}(a),
the chiralities $\tau =\pm 1$ are defined on the faces of the plaquettes
where the corresponding gauge invariant phase differences between two
nearest neighbor spins are $\varphi _{ij}\equiv \theta _{i}-\theta
_{j}+A_{ij}=\pm \pi /4$. Since all the choices of fully frustrated gauge
fields are physically equivalent, the gauge field given in Fig.~\ref%
{fig:phase}(a) are used throughout this paper. In the Coulomb gas language,
the $\tau _{p}=\pm 1$ chiralities can be viewed as $q_{p}=\pm \frac{1}{2}$
topological charges located at the centers of the plaquettes.

As the temperature increases, two kinds of topological excitations are
expected to disorder the system: (i) point-like defects as vortices or
anti-vortices which destroy the $U(1)$ phase coherence by flipping the signs
of the topological charges, (ii) linear defects as the domain walls
separating two ground states of different checkerboard patterns of
topological charges. Hence, the FFXY model is expected to have two kinds of
phase transitions associated with the formation of the quasi-long-range
order of the $U(1)$ spins and the long-range Ising order characterized by
chirality. Besides, the close interplay between different topological
excitations makes it difficult but interesting to explore the nature of the
transitions.

\section{Tensor network theory}

\subsection{Representations of partition function}

The partition function of a classical lattice model with local interactions
can be always represented as a contraction of tensor network on its original
lattice. The standard construction of the network starts from putting an
interaction matrix on each bond accounting for the Boltzmann weight. Then
the local tensors defined on the lattice sites are obtained by taking
suitable decompositions for the local bond matrices. Although this paradigm
has been proven a success in the studies of the classical XY model\cite%
{Yu2014,Vanderstraeten2019_2,Song2021}, it cannot be directly applied to the
fully frustrated case where the constraints of the ground-state local rules
should be imposed at the level of the local tensors\cite%
{Vanderstraeten2018,Vanhecke2021}.

To illustrate this point, we first derive the partition function of the FFXY
model following the standard approach. The partition function on the
original lattice is expressed as
\begin{equation}
Z=\prod_{i}\int \frac{\mathrm{d}\theta _{i}}{2\pi }\prod_{\Box
_{ijkl}}W(\theta _{i},\theta _{j})W^{\prime }(\theta _{j},\theta
_{k})W(\theta _{k},\theta _{l})W(\theta _{l},\theta _{i}),
\end{equation}%
where
\begin{equation*}
W(\theta _{i},\theta _{j})=\mathrm{e}^{\beta J\cos (\theta _{i}-\theta
_{j})},W^{\prime }(\theta _{i},\theta _{j})=\mathrm{e}^{-\beta J\cos (\theta
_{i}-\theta _{j})}
\end{equation*}%
can be viewed as the infinite interaction matrices with continuous $U(1)$
indices and $\beta =1/T$ is the inverse temperature. The partition function
is now cast into the tensor network representation as shown in Fig.~\ref%
{fig:tensor_err}(a), where the integrations $\int d\theta _{i}/2\pi $ is
denoted as red dots and the matrix indices take the same values at the joint
points.

\begin{figure}[tbp]
\centering
\includegraphics[width=\linewidth]{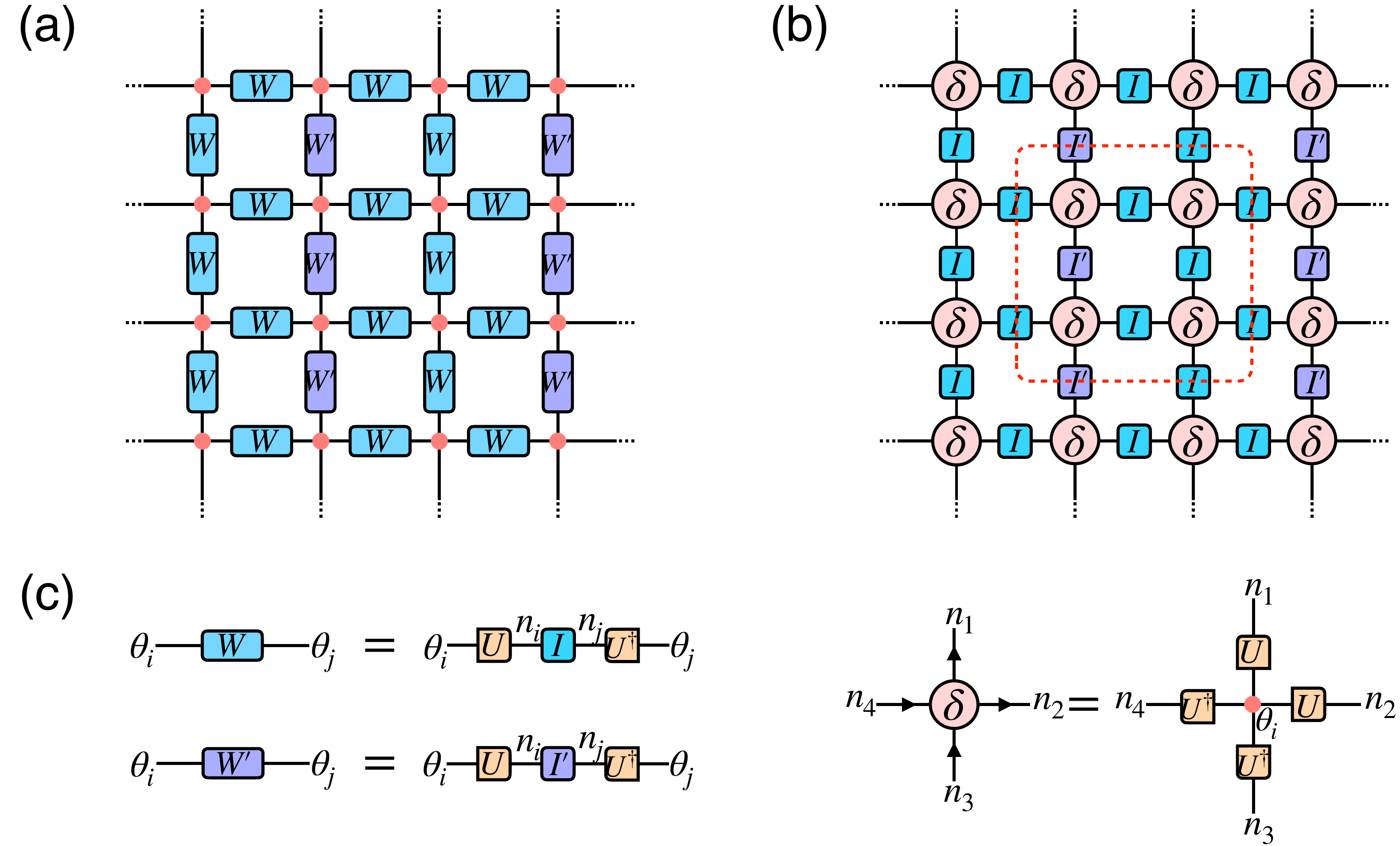}
\caption{ (a) Tensor network representation of the partition function with
interaction matrices on the links accounting for the Boltzmann weight. (b)
Tensor network representation of the partition function defined on the
original lattice. The translation invariant cluster is circled by the red
dotted line. (c) The eigenvalue decompositions of the interaction matrices
and the construction of delta tensor.}
\label{fig:tensor_err}
\end{figure}

To transform the local tensors into a discrete basis, we employ the
character decomposition for the Boltzmann factor,
\begin{equation}
\mathrm{e}^{x\cos \theta }=\sum_{n=-\infty }^{\infty }I_{n}(x)\mathrm{e}%
^{in\theta }
\end{equation}
where $I_{n}(x)$ are the modified Bessel functions of the first kind. The
eigenvalue decompositions are expressed as $U_{\theta ,n}=\mathrm{e}%
^{in\theta }$ and $I_{n}^{\prime n}\left( \beta J\right) =\left( -1\right)
^{n}I_{n}(\beta J)$, shown in Fig.~\ref{fig:tensor_err}(c). The integration
over all site-variables is now transformed into a product of independent
integrations of all plane waves. It is easy to integrate out the phase
degrees of freedom at each site
\begin{equation}
\int \frac{d\theta _{i}}{2\pi }U_{\theta _{i},n_{1}}U_{\theta
_{i},n_{1}}U_{\theta _{i},n_{3}}^{\ast }U_{\theta _{i},n_{4}}^{\ast }=\delta
_{n_{1}+n2}^{n_{3}+n_{4}}.  \label{eq:delta}
\end{equation}%
Then the conservation law of $U(1)$ charges has been encoded in the local $%
\delta $ tensors as the constraint: $\delta _{n_{1}+n2}^{n_{3}+n_{4}}\neq 0$
only if $n_{1}+n_{2}=n_{3}+n_{4}$. As a result, the $U(1)$ degrees of
freedom are transformed into the discrete bond indices $n$, represented as
links in the tensor network whose structure is depicted in Fig.~\ref%
{fig:tensor_err}(b).

The real challenge comes from the construction of the local tensors under
the ground-state local rules. For the classical XY model, to build the
translation invariant local tensors, we can simply split the diagonal
spectrum $I_{n}$ tensors and take a contraction of four $\sqrt{I_{n}}$
tensors connected to the $\delta $ tensors at the same site\cite%
{Yu2014,Vanderstraeten2019_2}. For the FFXY model with a checkerboard-like
ground state, the translation invariant unit is a $2\times 2$ plaquette. So
it is reasonable to enlarge the unit cell as a cluster consisting of $%
2\times 2$ tensors, as circled by the red dotted line in Fig.~\ref%
{fig:tensor_err}(b). However, we find the standard contraction algorithms
such as variational uniform matrix product state (VUMPS)\cite%
{Stauber2018,Vanderstraeten2019,Nietner2020} and corner transfer matrix
renormalization group (CTMRG)\cite{Nishino1996,Orus2009,Corboz2014} fail to
converge in such a construction of local tensors.

Two important issues are needed to be addressed in this construction. First,
from the perspectives of tessellation, the constraints for the phase
differences between two nearest neighbor spins are only imposed on the four
lattice sites within a cluster. Since two nearest neighbor clusters are
separated by an intermediate plaquette, the constraints between the spins
across the gutter is lost. Second, the linear transfer matrix composed by an
infinite row/column of local tensors under this construction is always
non-Hermitian. The key point is that the spectrum tensors $I_{n}^{\prime }$
carry a negative factor $(-1)^{n}$ which can never be divided into two
Hermitian adjoint partitions. Moreover, the negative factors cannot be
eliminated under any local transforms due to the \textquotedblleft odd
rules" induced by the gauge field\cite{Villain1977}.

\begin{figure}[tbp]
\centering
\includegraphics[width=\linewidth]{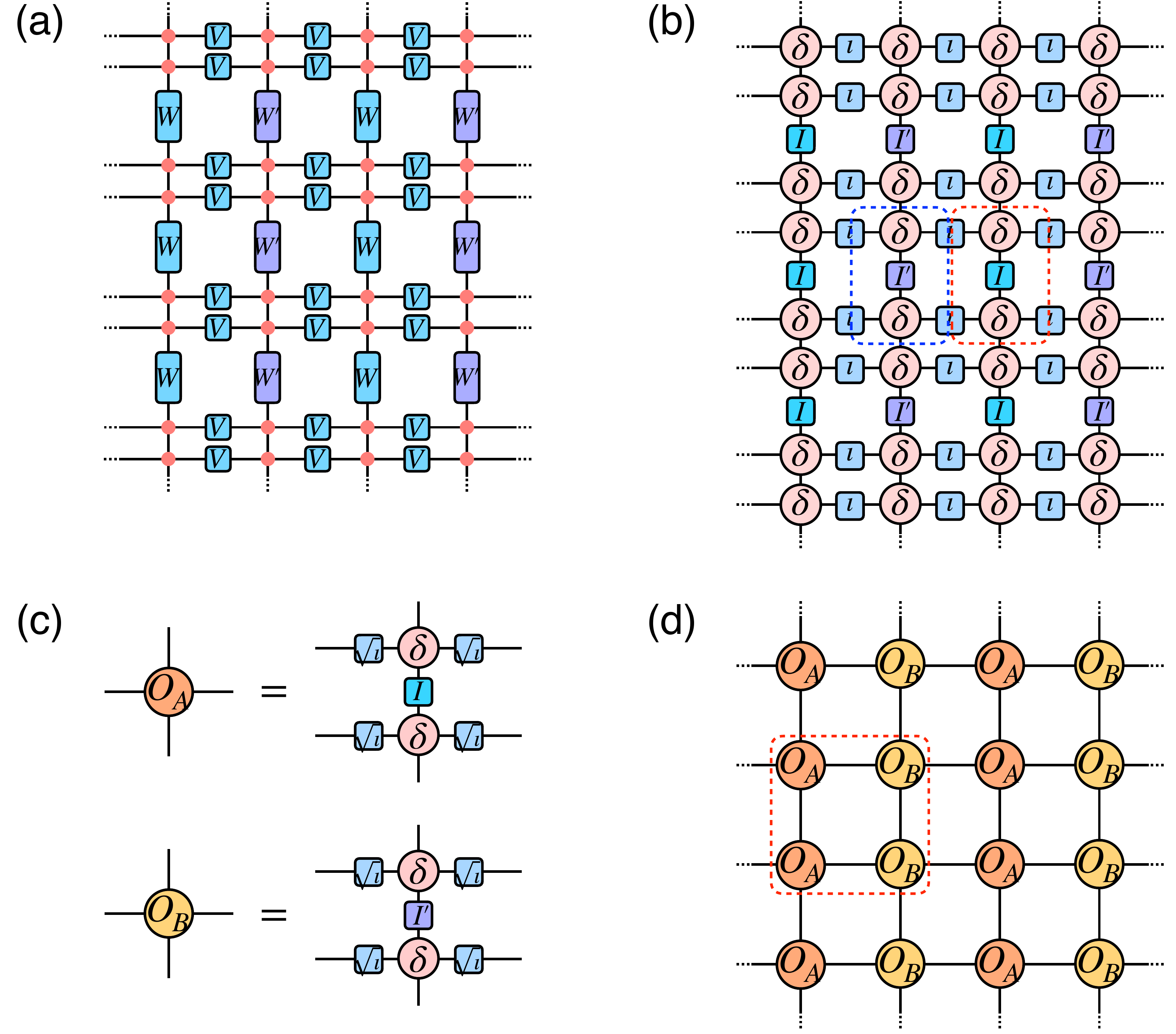}
\caption{ (a) Tensor network representation of the partition function with
vertically split $U(1)$ spins. (b) The transformation to discrete degrees of
freedom by integrating out $U(1)$ phase variables. (c) The construction of
the local $O_A$ and $O_B$ tensors. (d) The appropriate construction of the
tensor network for the partition function encoding the constraints.}
\label{fig:tensor_sp}
\end{figure}

In order to solve these problems, we propose a new construction method based
on the split of $U(1)$ spins on the lattice site. As shown in Fig.~\ref%
{fig:tensor_sp}(a), each lattice spin is vertically split into two
independent spins by using the relation
\begin{equation}
\int d\theta _{i}\,f(\theta _{i})=\iint d\theta _{i}d\theta _{i}^{\prime
}\,\delta (\theta _{i}-\theta _{i}^{\prime })f(\theta _{i})
\end{equation}%
where $f(\theta _{i})$ denotes the rest part of the partition function
associated to the lattice site $i$. And the corresponding interaction
matrices $W$ are also equally divided into two $V$ tensors as
\begin{equation}
V(\theta _{i},\theta _{j})=\mathrm{e}^{\frac{1}{2}\beta J\cos (\theta
_{i}-\theta _{j})}.
\end{equation}%
Then, we carry out eigenvalue decompositions on the $V$ tensor in the same
manner,
\begin{equation}
V(\theta _{i},\theta _{j})=\sum_{n}U_{\theta _{i},n}\,\iota _{n}(\beta
J)\,U_{n,\theta _{j}}^{\ast },
\end{equation}%
where $\iota _{n}(\beta J)=I_{n}(\beta J/2)$. The Dirac delta function
connecting two cloned spins can be decomposed by
\begin{equation}
\delta (\theta _{i},\theta _{i}^{\prime })=\frac{1}{2\pi }\sum_{n}U_{\theta
_{i},n}\,U_{n,\theta _{i}^{\prime }}^{\ast }.
\end{equation}%
Again the orthogonality of $U_{n,\theta }$ enables us to integrate out the
phase variables at each lattice site, and the enlarged tensor network is
displayed in Fig.~\ref{fig:tensor_sp}(b).

Now, we are able to walk around the obstacles by an appropriate construction
of the network. The minimum building blocks are the local tensors $O_{A}$
and $O_{B}$, whose inner structures are shown in Fig.~\ref{fig:tensor_sp}%
(c). The resulting tensor network $\mathcal{N}$ for the partition function
is depicted in Fig.~\ref{fig:tensor_sp}(d) as
\begin{equation}
Z=\text{\textrm{tTr}}\left( \prod_{O_{A},O_{B}\in \mathcal{N}%
}O_{A}O_{B}\right) ,  \label{eq:TN}
\end{equation}%
where \textquotedblleft tTr" denotes the tensor contraction over all
auxiliary links. Because the expansion coefficients decrease exponentially
in $\iota _{n}(x)$ with increasing $n$, an appropriate truncation can be
performed on the virtual indices of $O_{A}$ and $O_{B}$ tensors without loss
of accuracy. The constraints among four spins within a given plaquette is
ensured by choosing a cluster grouping $O_{A}$ and $O_{B}$ tensors. Although
the partition function is represented by the row-to-row transfer matrix
operator consisting of a single layer of alternating $O_{A}$ and $O_{B}$
tensors, it is only well-defined by even rows due to the non-trivial $%
2\times 2$ plaquette structure of the checkerboard ground state. Therefore,
the unit cluster should be composed by a double stack of $O_{A}$ and $O_{B}$
tensors as grouped by the red dotted line in Fig.~\ref{fig:tensor_sp}(d).
Ultimately, we obtain the right construction of the tensor network with the
linear transfer matrix consisting of the $2\times 2$ clusters. Such a
construction gives rise to the right partition function by realizing that
(i) all the constraints are preserved within the transfer matrix while the
spins across the gutter between two transfer matrices are indeed the same
spin, (ii) the transfer matrix is Hermitian as the splitting of the $%
I^{\prime }$ tensors is no longer needed.

Apart from the representation in the original lattice, there is another
approach to express the partition function as a tensor network on the dual
lattice with automatically encoded local constraints. For a model with
discrete degrees of freedom, the dual construction can always be performed
by splitting of the model Hamiltonian on a shared bond, and the local
tensors are defined on the plaquette centers by grouping the split bonds
which are connected by Kronecker delta functions. However, this strategy
cannot be simply extended for the case of the continuous degrees of freedom.
When we split each bond around the plaquette, there will be integrals of
loops of Dirac delta functions, which is not well-defined mathematically.
That is why we can only split the $W$ tensors in Fig.~\ref{fig:tensor_err}%
(a) horizontally. We also notice that the duality transformation of the FFXY
model onto the dual height model cannot give the appropriate partition
function, as a finite truncation on the height can mess up the Boltzmann
weights. Therefore, the construction of the tensor network in the dual space
remains an open problem.

\subsection{Multisite VUMPS algorithm}

Within the framework of tensor network, the fundamental object for the
calculation of the partition function is the transfer operator composed of
two infinite rows of alternating $O_{A}$ and $O_{B}$ tensors
\begin{equation}
\mathcal{T}(\beta )=\mathrm{{tTr}\,\left[
\begin{array}{c}
{{\cdots O_{A}O_{B}O_{A}O_{B}\cdots }} \\
{{\cdots O_{A}^{\prime }O_{B}^{\prime }O_{A}^{\prime }O_{B}^{\prime }\cdots }%
}%
\end{array}%
\right] ,}
\end{equation}%
where the prime symbols are just a mark to distinguish the second row from
the first row. This operator can be regarded as the matrix product operator
(MPO) for the 1D quantum spin chain, whose logarithmic form can be mapped to
a 1D quantum system with complicated spin-spin interactions. In this way,
the correspondence between the finite temperature 2D statistical model and
the 1D quantum model at zero temperature is established.

\begin{figure}[tbp]
\centering
\includegraphics[width=\linewidth]{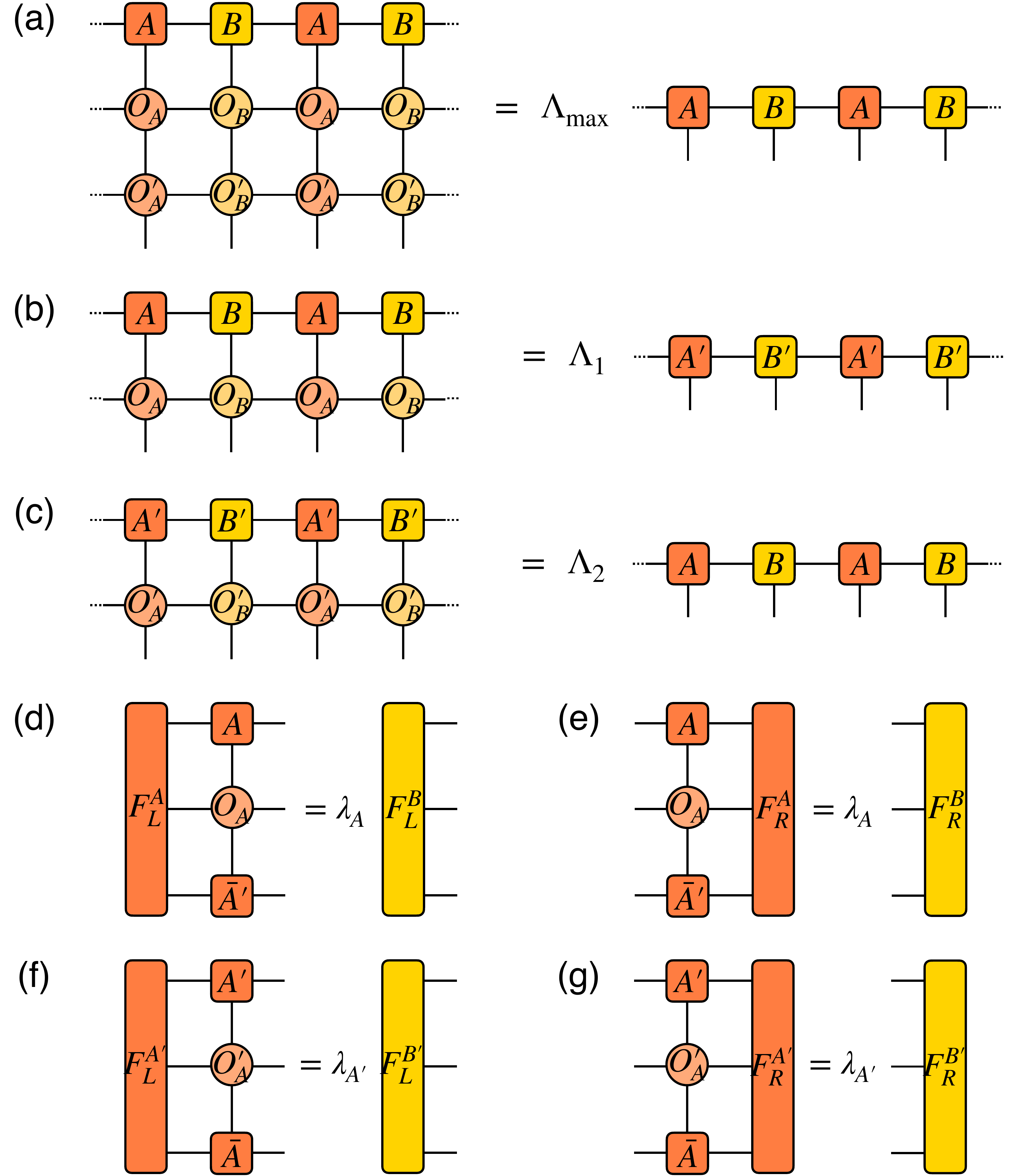}
\caption{ (a) The eigen equation for the fixed-point MPS $|\Psi(A, B)\rangle$
of the transfer operator $\mathcal{T}(\protect\beta)$. (b) and (c) Two
smaller linear equations for the first and second row as a decomposition of
the fixed-point eigen equation. (d)-(g) Eigen equations for the left and
right fixed points of the corresponding channel operators related to $O_A$
and $O_A^{\prime }$. }
\label{fig:vumps}
\end{figure}

As sketched in Fig.~\ref{fig:vumps}(a), the value for the partition function
is determined by the dominant eigenvalues of the transfer matrix
\begin{equation}
\mathcal{T}(\beta )|\Psi (A,B)\rangle =\Lambda _{\max }|\Psi (A,B)\rangle ,
\label{eq:TM}
\end{equation}%
where $|\Psi (A,B)\rangle $ is the leading eigenvector represented by matrix
product states (MPS) made up of two-site unit cell of local $A$ and $B$
tensors\cite{Stauber2018}. This fixed-point equation can be accurately
solved by the multiple lattice-site VUMPS algorithm\cite{Nietner2020}, which
provides an efficient variational scheme to approximate the largest
eigenvector $|\Psi (A,B)\rangle $. The precision of this approximation is
controlled by the auxiliary bond dimension $D$ of local $A$ and $B$ tensors.
Instead of grouping the local tensors of a cluster into a trivial unit cell
at the cost of growing the bond dimension of the MPO exponentially, the
multisite VUMPS algorithm brings about a significant speed-up with a
computational complexity only scaling linearly with the size of the
multisite cluster.

The multisite algorithm starts from decomposing the big eigen equation into
two smaller linear equations as shown in Fig.~\ref{fig:vumps}(b) and (c):
\begin{align}
T_{1}(\beta )|\Psi (A,B)\rangle & =\Lambda _{1}\Psi (A^{\prime },B^{\prime
})\rangle ,  \notag \\
T_{2}(\beta )|\Psi (A^{\prime },B^{\prime })\rangle & =\Lambda _{2}|\Psi
(A,B)\rangle ,
\end{align}%
where $T_{1}(\beta )$ and $T_{2}(\beta )$ correspond to the first and second
row of the blocked transfer matrix $\mathcal{T}(\beta )$, whose eigenvalue
is a combination as $\Lambda _{\max }=\Lambda _{1}\Lambda _{2}$. In
practice, the linear equations are transformed into a set of optimization
problems
\begin{align*}
\Lambda _{1}& =\max_{A,B,A^{\prime },B^{\prime }}\langle \Psi
(A,B)|T_{1}(\beta )|\Psi (\bar{A^{\prime }},\bar{B^{\prime }})\rangle , \\
\Lambda _{2}& =\max_{A,B,A^{\prime },B^{\prime }}\langle \Psi (A^{\prime
},B^{\prime })|T_{1}(\beta )|\Psi (\bar{A},\bar{B})\rangle ,
\end{align*}%
which can be solved efficiently by applying the VUMPS algorithm\cite%
{Stauber2018,Vanderstraeten2019} based on the tangent space projections in
iteration.

One of the key steps in the iteration of VUMPS method is the calculation of
the leading left- and right-eigenvectors of the channel operators. The
channel operators have a sandwich structure composed of two local tensors of
fixed point MPS and the middle four-leg local tensor. The channel operator
related to $A$ tensor is defined by
\begin{equation}
\mathbb{T}_{X}^{A}=\sum_{i,j}\bar{A}^{i}\otimes X^{i,j}\otimes \bar{%
A^{\prime }}^{j},
\end{equation}%
and other channel operators are defined in the same way. For the $2\times 2$
unit cell, there are four set of eigen equations for the left- and
right-fixed points of the corresponding channel operators. Fig.~\ref%
{fig:vumps}(d)-(g) displays the eigen equations related to $O_{A}$ and $%
O_{A^{\prime }}$
\begin{align}
\left\langle F_{L}^{A}\right\vert \mathbb{T}_{O_{A}}^{A}=\lambda
_{A}\left\langle F_{L}^{B}\right\vert ,\ \ & \mathbb{T}_{O_{A}}^{A}\left%
\vert F_{R}^{A}\right\rangle =\lambda _{A}\left\vert F_{R}^{B}\right\rangle ,
\notag \\
\langle F_{L}^{A^{\prime }}|\mathbb{T}_{O_{A}^{\prime }}^{A^{\prime
}}=\lambda _{A^{\prime }}\langle F_{L}^{B^{\prime }}|,\ \ & \mathbb{T}%
_{O_{A}^{\prime }}^{A^{\prime }}|F_{R}^{A^{\prime }}\rangle =\lambda
_{A^{\prime }}|F_{R}^{B^{\prime }}\rangle ,
\end{align}%
and the same method are applied to $O_{B}$ and $O_{B^{\prime }}$. The above
equations only map two fixed points in the same row to each other without
directly giving an eigenvalue problem. In the same spirit as the solvers for
fixed-point MPS, these equations are iteratively applied until a given
convergence criterion is reached.

\subsection{Calculations of the physical quantities}

Once the fixed point MPS is achieved, various physical quantities can be
accurately calculated in the tensor-network language. The entanglement
properties can be detected via the Schmidt decomposition of $|\Psi(A,
B)\rangle$ which bipartites the relevant 1D quantum state of the MPO, and
the entanglement entropy\cite{Vidal2003} is determined directly from the
singular values $s_\alpha$ as
\begin{equation}
S_E = -\sum_{\alpha=1}^{D}s_\alpha^2\ln s_{\alpha}^2,  \label{eq:ee}
\end{equation}
in correspondence to the quantum entanglement measure for a many-body
quantum system.

\begin{figure}[tbp]
\centering
\includegraphics[width=\linewidth]{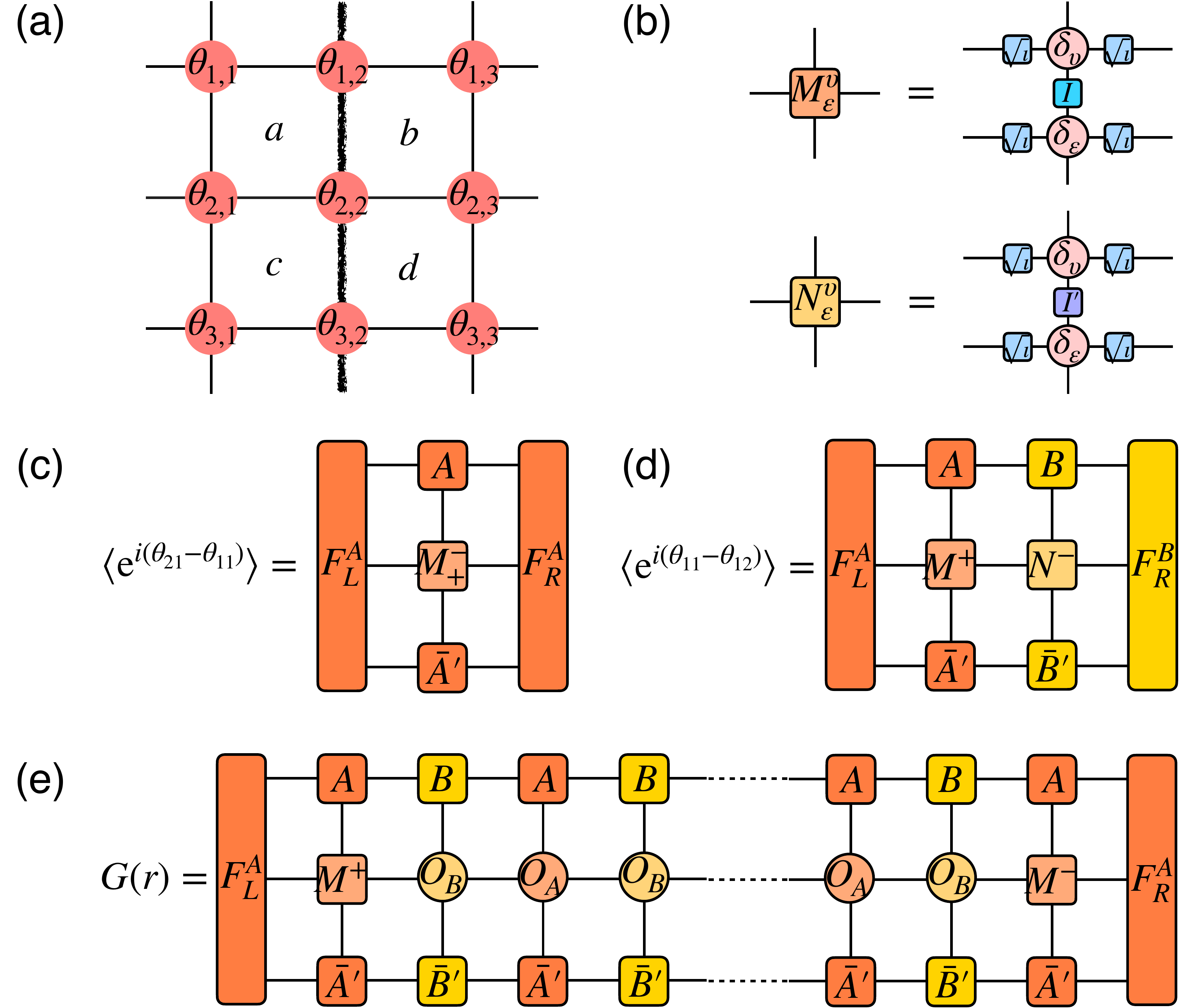}
\caption{ (a) A unit cell of $2\times2$ plaquettes with $9$ spins at
vertices. (b) The construction of the impurity tensors from imbalanced delta
tensors. (c) and (d) Evaluations of the nearest-neighbor two-angle
observables living on the vertical and horizontal bonds. (e) Two-point
correlation function represented by contracting a sequence of channel
operators. }
\label{fig:op_clfn}
\end{figure}

Local observable can be evaluated by inserting the corresponding impurity
tensors into the original tensor network for the partition function. We can
squeeze the whole network into an infinite chain of channel operators by
sequentially pulling the MPS fixed points through the network from top and
bottom. Then a further contraction is performed by the left and right fixed
points of the channel operators. For instance, the expectation value of the
local chirality at a plaquette $p$ is defined as
\begin{equation}
\tau _{p}=\frac{1}{2\sqrt{2}}\sum_{\langle i,j\rangle \in \Box
_{p}}\left\langle \sin (\theta _{i}-\theta _{j}+A_{ij})\right\rangle ,
\label{eq:tau}
\end{equation}%
where the sum runs over the four bonds around the plaquette anti-clockwise,
corresponding to four pairs of nearest-neighbor two-angle observable. For
the FFXY model with checkerboard like order of the chirality, it is
necessary to pick out a unit cell of $2\times 2$ plaquette as shown in Fig.~%
\ref{fig:op_clfn}(a), where the sub-plaquettes are labeled with $a$, $b$, $c$
and $d$. The number of two-angle observable needed to evaluate the local
chirality of the sub-lattices can be reduced from twelve to eight due to the
transitional symmetry of the unit cell. It is easy to check the identities
between the observable at boundaries like
\begin{eqnarray}
\langle \mathrm{e}^{i(\theta _{1,1}-\theta _{1,2})}\rangle  &=&\langle
\mathrm{e}^{i(\theta _{3,1}-\theta _{3,2})}\rangle   \notag \\
&=&\frac{1}{Z}\prod_{i}\frac{d\theta _{i}}{2\pi }\mathrm{e}^{-\beta
E(\{\theta _{i}\})}\mathrm{e}^{i\theta _{1,1}}\mathrm{e}^{-i\theta _{1,2}},
\end{eqnarray}%
where $E(\{\theta _{i}\})$ is the energy for a given spin configuration.
Compared to the orthogonal relation of \eqref{eq:delta}, $\mathrm{e}%
^{i\theta _{1,1}}$ and $\mathrm{e}^{-i\theta _{1,2}}$ in the second row
simply change the corresponding delta tensors in the original tensor network
for the partition function in Fig.~\ref{fig:tensor_sp}(b) into
\begin{equation}
\delta _{+}=\delta _{n_{1}+n_{2}+1}^{n_{3}+n_{4}},\quad \delta _{-}=\delta
_{n_{1}+n_{2}}^{n_{3}+n_{4}+1},
\end{equation}%
which introduce the impurity tensors $M^{+}$ and $N^{-}$ containing these
imbalanced delta tensors into the tensor network of Fig.~\ref{fig:tensor_sp}%
(d). The structure of the impurity tensors are displayed in Fig.~\ref%
{fig:op_clfn}(b) as $M_{\varepsilon }^{\upsilon }$ and $N_{\varepsilon
}^{\upsilon }$ in replacement of $O_{A}$ and $O_{B}$, respectively, where $%
\varepsilon ,\upsilon =\pm $ are in consistent with $\delta _{\pm }$. Here
the superscript and subscript in $M$ and $N$ are omitted when there is a
normal delta tensor defined by \eqref{eq:delta}. With the help of the
fixed-points of the channel operators, it is straightforward to get the
two-angle observable sharing the vertical bonds:
\begin{equation}
\langle \mathrm{e}^{i(\theta _{2,1}-\theta _{1,1})}\rangle =\langle
F_{L}^{A}|\mathbb{T}_{M_{+}^{-}}^{A}|F_{R}^{A}\rangle ,
\end{equation}%
and those living on the horizontal bonds:
\begin{equation}
\langle \mathrm{e}^{i(\theta _{1,1}-\theta _{1,2})}\rangle =\langle
F_{L}^{A}|\mathbb{T}_{M^{+}}^{A}\mathbb{T}_{N^{-}}^{B}|F_{R}^{B}\rangle ,
\end{equation}%
as graphically depicted in Fig.~\ref{fig:op_clfn}(c) and (d). Finally, we
deduce the local chiralities at four sub-plaquettes from the imaginary part
of theses two-angle observable and the internal energy per site can be
obtained readily from the real part as
\begin{equation}
u=-\frac{J}{2}\sum_{\langle i,j\rangle \in \Box }\left\langle \cos (\theta
_{i}-\theta _{j}+A_{ij})\right\rangle .  \label{eq:u}
\end{equation}

Moreover, the two-point correlation function between local observable is
defined by $G(r)=\langle h(\theta _{j})h(\theta _{j+r})\rangle $, which can
be evaluated by inserting two local impurity tensors into the original
tensor network. The corresponding impurity tensors are constructed in the
same way by altering the Kronecker delta tensors:
\begin{equation}
\delta _{n_{1},n_{2}}^{n_{3},n_{4}}\rightarrow \int \frac{d\theta }{2\pi }\,%
\mathrm{e}^{i(n_{1}+n_{2}-n_{3}-n_{4})}h(\theta ).
\end{equation}%
For the spin-spin correlation function, as shown in Fig.~\ref{fig:op_clfn}%
(e), the evaluation of $G(r)=\langle \mathrm{e}^{i(\theta _{j}-\theta
_{j+r})}\rangle $ is reduced to a trace of a row of channel operators
containing two impurity tensors $M^{+}$ and $N^{-}$
\begin{equation}
G(r)=\langle F_{L}^{A}|\mathbb{T}_{M^{+}}^{A}\underbrace{\mathbb{T}_{O}^{B}%
\mathbb{T}_{O}^{A}\mathbb{T}_{O}^{B}\cdots \mathbb{T}_{O}^{A}\mathbb{T}%
_{O}^{B}}_{r-1}\mathbb{T}_{M^{-}}^{A}|F_{R}^{A}\rangle ,  \label{eq:G}
\end{equation}%
where the left and right leading eigenvectors of the channel operators are
employed.

\section{Numerical Results}

Most of the previous studies determine the transition temperature according
to some kind of order parameter like the magnetization or Binder cumulant.
These order parameters are good criteria to identify the critical
temperatures relevant to $U(1)$ or $Z_{2}$ phase transition when several
transitions are apart from each other with sufficient distance separation.
However, for the case of FFXY model, it is hard to tell apart two kinds of
mutually close transitions from these quantities because either the $U(1)$
or $Z_{2}$ transition temperature are obtained from an average with some
degrees of uncertainty and the estimated temperatures may mix up with each
other. Here, we propose that the entanglement entropy shall shed a new light
to overcome the difficulties in deciding whether there are two distinct
transitions in the FFXY model. The entanglement entropy of the fixed-point
MPS for the 1D quantum transfer operator exhibits singularity at the
critical temperatures which offers a sharp criteria to accurately determine
all possible phase transitions, especially for systems possessing $%
U(1)\times Z_{2}$ symmetry\cite{Song2021,Song2021_2}.

\begin{figure}[tbp]
\centering
\includegraphics[width=\linewidth]{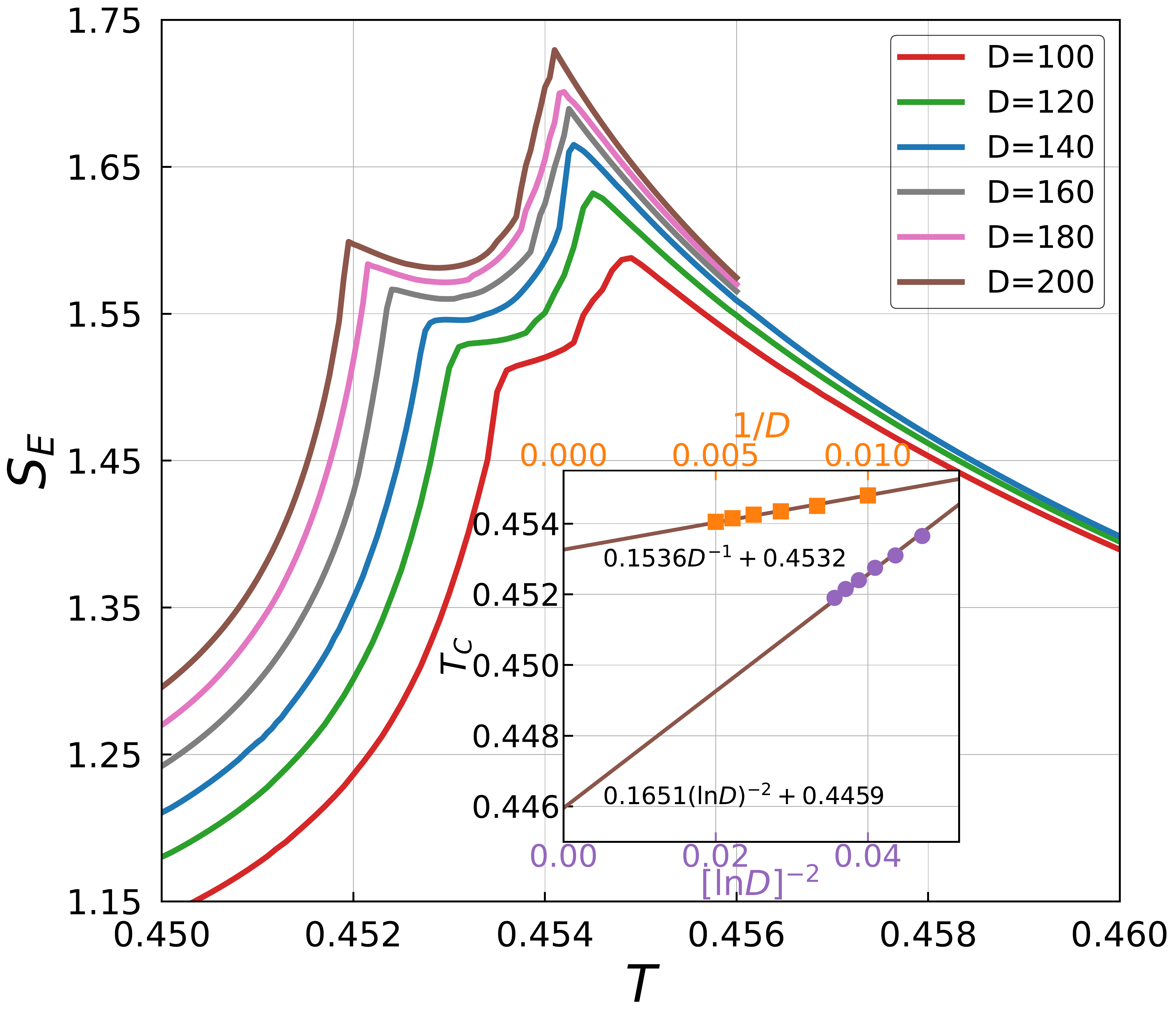}
\caption{ The entanglement entropy as a function of temperature develops two
singularities at $T_{c1}$ and $T_{c2}$ indicating two phase transitions for
different MPS bond dimensions. Inset: The singularity temperatures $T_{c1}$
and $T_{c2}$ of the entanglement entropy fitted for MPS bond dimensions from
$100$ to $200$ with an interval of $20$. The lower transition temperature $%
T_{c1}$ varies linearly on $(\ln D)^{-2}$, while the higher transition
temperature $T_{c2}$ has a linear variance with $1/D$.}
\label{fig:es}
\end{figure}

As shown in Fig.~\ref{fig:es}, the entanglement entropy $S_{E}$ develops two
sharp singularity at two critical temperatures $T_{c1}$ and $T_{c2}$, which
strongly indicates the existence of two phase transitions at two different
temperatures. As the singularity positions vary with the MPS bond dimension $%
D$, the critical temperatures $T_{c1}$ and $T_{c2}$ can be determined
precisely by extrapolating the bond dimension $D$ to infinite. Moreover, we
find that the critical temperatures $T_{c1}$ and $T_{c2}$ exhibit different
scaling behavior in the linear extrapolation, implying that that the two
phase transitions belong to different kinds of universality class. The inset
of Fig.~\ref{fig:es} displays how the critical temperatures, $T_{c1}$ and $%
T_{c2}$, vary with the MPS bond dimensions. The lower transition temperature
$T_{c1}$ varies linearly on the inverse square of the logarithm of the bond
dimension, while the higher transition temperature $T_{c2}$ has a linear
variance with the inverse bond dimension. From the linear extrapolation, the
critical temperatures are estimated to be $T_{c1}\simeq 0.4459$ and $%
T_{c2}\simeq 0.4532$, which agrees well with the previous Monte Carlo
simulations\cite{Olsson1995,Okumura2011}.

Actually, the different scaling behavior stems from the different critical
behavior of the BKT and 2D Ising transitions. The BKT transition for $T>T_C$
is characterized by the exponentially diverging correlation length
\begin{equation}
\xi_{BKT}(T)\propto\exp(b/\sqrt{T-T_C}), \notag
\end{equation}
while the 2D Ising transition is featured by
$$\xi_{Ising}(T)\propto1/|T-T_C|.$$
Since the bond dimensions of the fixed-point MPS can be regarded as a finite cutoff on
the diverging correlation length, it is reasonable to use the $(\ln D)^{-2}$
and the $1/D$ scaling for the extrapolation of the critical temperatures $T_{c1}$
and $T_{c2}$, respectively. Besides, the separation between $T_{c1}$ and $T_{c2}$ gets
larger as the bond dimension increases, which indicates that large bond dimensions are
necessary to clarify the nature of the transitions.

\begin{figure}[tbp]
\centering
\includegraphics[width=\linewidth]{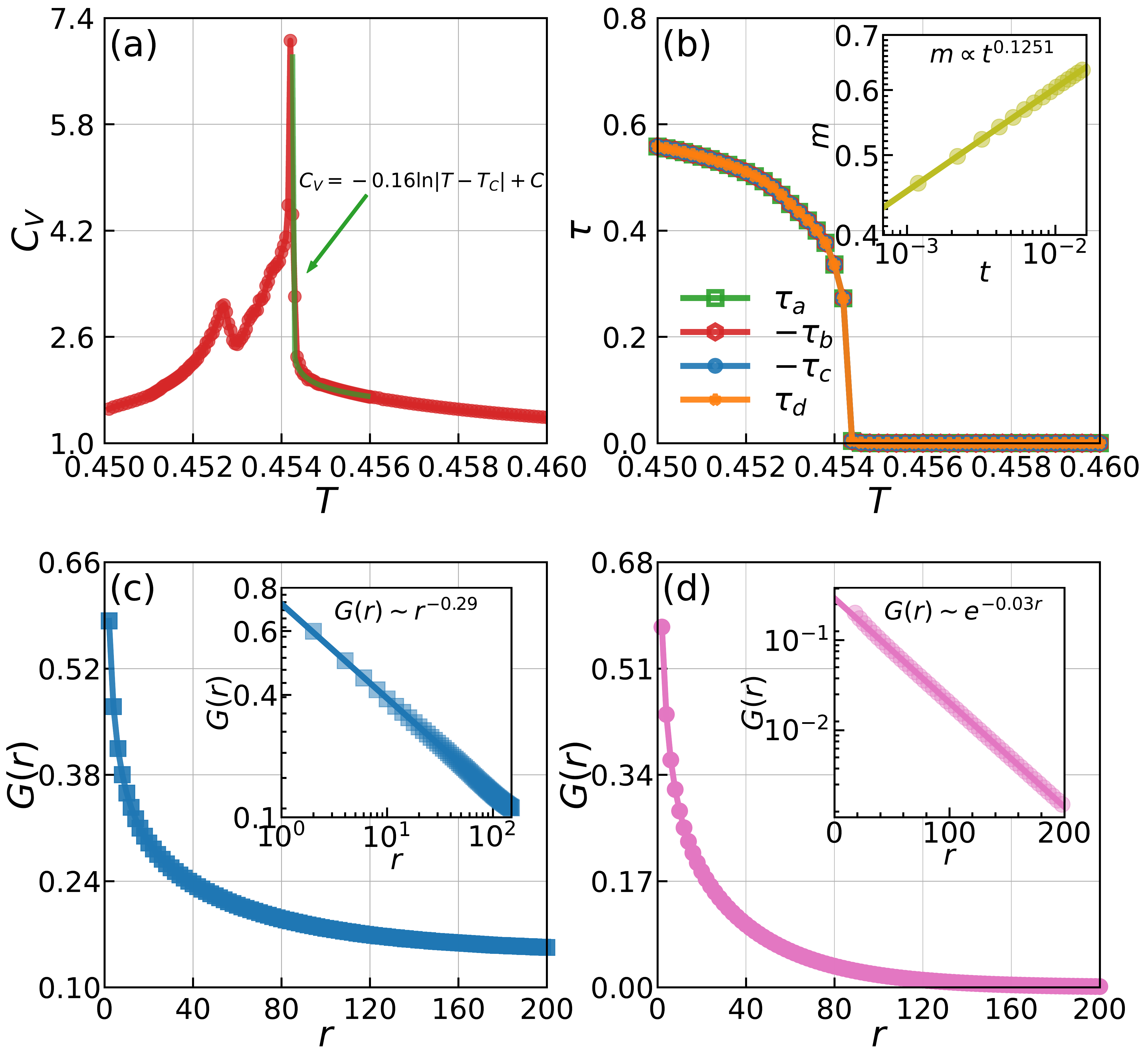}
\caption{ (a) The specific heat develops a small bump and sharp peak at $%
T_{c1}$ and $T_{c2}$, respectively. The green line is a fitting curve of the logarithmic divergence for $T>T_{c2}$.
(b) The establishment of the
checkerboard like pattern of local chiralities below $T_{c2}$. Inset is the
fitting of the critical exponent of the staggered magnetization. (c) The
spin-spin correlation function displays power law behavior at $T=0.450$
below $T_{c1}$. (d) The spin-spin correlation function decays exponentially
at $T=0.454$ at the intermediate temperature between $T_{c1}$ and $T_{c2}$.}
\label{fig:tau_cv}
\end{figure}

In order to gain insight into the essential physics of different phase
transitions, we investigate the thermodynamic properties. We begin with the
specific heat which can be derived directly from $C_{V}=du/dT$, where $u$ is
the internal energy obtained by the contraction of the tensor network with
the nearest-neighbor impurity tensors as introduced in \eqref{eq:u}. As
shown in Fig.~\ref{fig:tau_cv}(a), the specific heat exhibits a logarithmic
divergence at $T>T_{c2}$ but a small bump around $T_{c1}$. The logarithmic
specific heat at the higher temperature side indicates the occurrence of a
second-order phase transition such as the 2D Ising phase transition, while
the small bump at the lower temperature can be regarded as a higher order
continuous phase transition like the BKT transition.
The specific heat curve is inadequate for a logarithmic fitting at the lower temperature side of $T_{c2}$ because two different transitions are still closed to each other.

It is natural to expect that the logarithmic peak of the specific heat is
related to the breaking of the chiral order as the FFXY has a checkerboard
ground state with $Z_{2}$ symmetry. We thus check this long-range ordering
of the local chirality $\tau $ at the sub-plaquettes defined by %
\eqref{eq:tau} within a transition invariant unit cell depicted in Fig.~\ref%
{fig:op_clfn}(a). As shown in Fig.~\ref{fig:tau_cv}(b), the expectation
values of the local chirality are finite below $T_{c2}$, indicating the
formation of the long-range $Z_{2}$ order. In addition, we find a perfect
agreement for the four chiralities as $\tau _{a}=-\tau _{b}=-\tau _{c}=\tau
_{d}$, which is a clear evidence for the emergence of the checkerboard
pattern through the phase transition at $T_{c2}$. The checkerboard like
order of the chirality can be characterized by a staggered magnetization
defined as
\begin{equation}
m=\frac{1}{N}\sum_{\Box _{p}}(-1)^{x_{p}+y_{p}}\tau _{p},
\end{equation}%
where $\tau _{p}$ is the local chirality at the center of the plaquette
located at position $(x_{p},y_{p})$. As the temperature approaching $T_{c2}$
from below, we find that $m$ vanishes continuously as $m\sim t^{\beta }$
with $t\equiv (T_{c2}-T)/T_{c2}$, where $\beta \simeq 1/8$ is the critical
exponents characteristic of the 2D Ising model. The linear fitting of $\ln
m\propto \beta \ln t$ is depicted in the inset of Fig.~\ref{fig:tau_cv}(b).
Therefore, there is a convincing evidence that the transition at $T_{c2}$
belongs to the 2D Ising universality class.

From the analysis of the specific heat, we are aware of the transition at $%
T_{c1}$ related to a spin ordering, which is completely different from the
second order phase transitions. To further explore the nature of the phase
transition, we calculate the spin-spin correlation function
\begin{equation*}
G(r)=\langle \psi ^{\ast }(r)\psi (0)\rangle \propto \langle \cos (\theta
_{0}-\theta _{r})\rangle
\end{equation*}
within the tensor network framework by \eqref{eq:G}. As displayed in Fig.~%
\ref{fig:tau_cv}(c), the spin-spin correlation function exhibits an
algebraic behavior below $T_{c1}$, implying the vortex-antivortex bindings
in the spin configuration. However, for the temperature above $T_{c1}$, $%
G(r) $ decays exponentially, indicating the destruction of phase coherence
between vortex pairs. Fig.~\ref{fig:tau_cv}(d) shows the exponential
behavior of $G(r)$ at an intermediate temperature between $T_{c1}$ and $%
T_{c2}$. Hence, the change in the behavior of the correlation function at $%
T_{c1}$ turns out to be in the universality class of the BKT transition.

Finally, the whole phase structure is summarized in Fig.~\ref{fig:phase}(b).
The FFXY model has two very close but separate phase transitions, with
transition temperature $T_{BKT}<T_{Ising}$. The transition at $T_{Ising}$
belongs to the usual 2D Ising universality class, while the transition at $%
T_{BKT}$ belongs to the BKT universality class. As the system cools down,
the $Z_{2}$ symmetry is first broken at $T_{Ising}$ characterized by the
formation of checkerboard like long-range order of chiralities, and then the
BKT transition occurs at a lower temperature $T_{BKT}$ featured by the
algebraic correlation between vortex-antivortex pairs.

\section{Conclusion}

We have proposed a general framework to solve the 2D FFXY model. The
important aspect is to encode the ground-state local rules induced by
frustrations into the local tensors of the partition function. Then the
partition function is written in terms of a product of 1D transfer matrix
operator, whose eigen-equation is solved by an algorithm of matrix product
states rigorously. The singularity of the entanglement entropy for this 1D
quantum analogue provides a stringent criterion to determine various phase
transitions without identifying any order parameter a prior. Certainly the
present methods provide a promising route to solve other frustrated lattice
models with continuous degrees of freedom in 2D.

The main result of our tensor network theory is that, higher than the BKT
phase transition, a chiral ordered phase with spontaneously broken
$Z_2$ symmetry has been confirmed, where the spin-spin phase coherence is absent but the phase differences of the spins on
two nearest neighbor sites have a nontrivial value different from $0$ or $\pi$. In contrast to the conventional
situations where the Ising transition happens at a lower temperature below the BKT transition, the ordering of the spins in the FFXY model is nontrivial. It is highly interesting that
the chiral ordered intermediate phase may be related to some unconventional superconductivity in the absence of condensed Cooper pairs.

Furthemore, another phase transition associated with the unbinding of kink pairs on domain walls in the FFXY model\cite{Korshunov2002} was proposed to support the existence of two separate bulk transitions $T_{BKT}<T_{Ising}$, which happens at a lower temperature $T_{kink}<T_{BKT}$. The kinks at the corners of domain walls behave as fractional vortices with the topological charge $\pm\frac{1}{4}$. However, the verification of such transition is only realized under a special boundary condition where an infinite domain wall is ensured\cite{Olsson2005}. We believe that our tensor network approach may provide a promising way for further detailed investigations of such kink-antikink unbinding transition.

\textbf{Acknowledgments} The authors are very grateful to Qi Zhang for his
stimulating discussions. The research is supported by the National Key
Research and Development Program of MOST of China (2017YFA0302902).

\bibliography{reference}

\end{document}